\documentclass[graybox]{svmult}

\usepackage{mathptmx}       
\usepackage{helvet}         
\usepackage{courier}        
\usepackage{type1cm}        
\usepackage{makeidx}         
\usepackage{graphicx}        
\usepackage{multicol}        
\usepackage[bottom]{footmisc}

\makeindex             

\begin{document}

\title*{Magnetic Fields}
\author{Markus Sch\"oller and Swetlana Hubrig}
\institute{Markus Sch\"oller \at European Southern Observatory, Karl-Schwarzschild-Str.~2, 85748 Garching, Germany, \email{mschoell@eso.org}
\and Swetlana Hubrig \at Leibniz-Institut f\"ur Astrophysik, An der Sternwarte~16, 14482 Potsdam, Germany \email{shubrig@aip.de}}
\maketitle

\abstract{
In this chapter, we give a brief introduction into the use of the Zeeman
effect in astronomy and the general detection of magnetic fields in stars,
concentrating on the use of FORS\,2 for longitudinal magnetic field measurements.
}

\section{Magnetic fields in stars}
\label{sect:MF_intro}

Magnetic fields are found everywhere in the cosmos, from very weak fields in the
interstellar medium to the fields of magnetars, which are many orders of magnitude
stronger than any field ever generated by a human.
It is probably true that every star has a magnetic field.
When reading in the literature that a star is magnetic, this usually means that it
is strongly magnetic, or even clearer, its magnetic field is strong enough to
be detected.
To get an idea of the strength of the magnetic fields in different kind of stars, we
give a brief overview here:

\begin{itemize}
\item{Vega}: $<1$\,G
\item{Sun}: $0.5-4$\,G, in sunspots $2-5$\,kG
\item{Babcock’s (Ap) star}: 34\,kG
\item{White dwarfs}: $10^3-10^9$\,G
\item{Neutron stars and magnetars}: $10^9-10^{15}$\,G
\end{itemize}

Magnetic fields in astronomy are typically measured in Gauss, where 1\,G
corresponds to 0.1\,mT.
Technically generated magnetic fields on earth are on the order of tens of Tesla
for permanent fields, or tens of kT for short-lived fields, which is still six to
seven orders of magnitude smaller than the strongest fields in the neutron stars.

Generally, astronomers have a love-hate relationship with magnetic fields. 
They are hated, since they make everything complicated, e.g.\ breaking the symmetries
in models.
On the other hand, magnetic fields are loved, because they are used as potential culprits
for differences between theory and observation.

Magnetic fields in stars may come from two different sources.
They might be of fossil origin, which implies that the magnetic flux $BR^2$
from the cloud they were born from was at least in part conserved during the cloud's collapse.
The magnetic flux is
the product of the magnetic field strength $B$ and the square of the radius $R$.
While the magnetic flux levels in the parent cloud are very low,
the resulting magnetic fields in the star would be enormous due
to the huge reduction in $R$ during the contraction before the star's birth and thus, an
efficient method is needed to lose the excessive magnetic flux.
Another source for the magnetic field could be a dynamo process, which
permanently regenerates the magnetic field using a seed field.
Dynamo action leads to the Sun's magnetic field and the solar cycle.

Magnetic fields are responsible for various processes in stars.
They can dominate the accretion process in pre-main sequence stars,
they are responsible for stellar activity like spots or flares, 
they can lead to chemical peculiarities,
they heat the corona, which then produces X-rays,
they brake stellar rotation and thus slow stars down,
and they accelerate cosmic ray particles in neutron stars and
are responsible for the pulses in the pulsars, a good example for oblique rotators.

\section{The Zeeman effect}
\label{sect:MF_zeeman}

\begin{figure}[t]
\centering
\includegraphics[width=0.45\textwidth, angle=270]{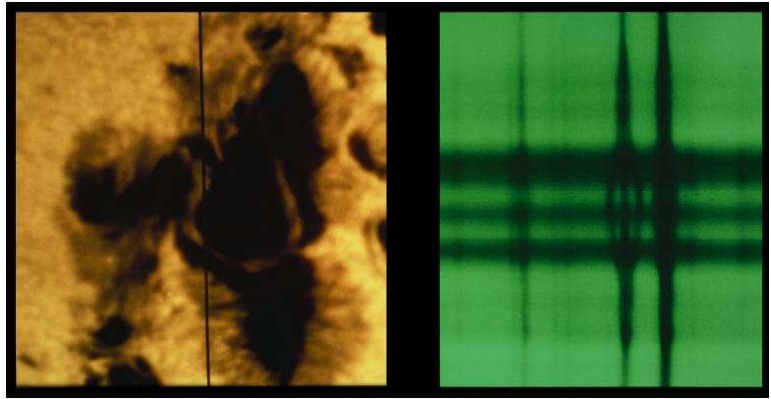}
\caption{
A complex sunspot picture, taken at 15:30 hours UT on 1974 July 4.
The vertical black line on the white light image (left) indicates the location
of the slit for the spectrograph that took the spectrum, shown on the right.
The division of one spectral line into three parts is a clear demonstration of the Zeeman effect.
In fact, the Zeeman splitting of this line, at 5250.2\,\AA{} and coming from iron,
indicates a record field strength of 4130\,G.
This picture was taken at the McMath-Pierce Solar Facility on Kitt Peak.
 -- Credit: NOAO/AURA/NSF.
}
\label{fig:sunspot}
\end{figure}

The Zeeman effect describes the splitting of a spectral line into several components
in the presence of a static magnetic field.
The distance between the Zeeman sub-levels is a function of the magnetic field $B$,
linear for weak fields.
It is proportional to the square of the wavelength, $\lambda^2$.
For strong magnetic fields, the splitting is governed by the Paschen-Back effect.
Transitions responsible for the spectral lines follow the selection rule that allows the difference of
the magnetic moment $\Delta{}m_J$ to assume only values of 0 (the $\pi$ component) or $\pm1$ (the $\sigma$ components).
In a magnetic field parallel to the line of sight,
the $\pi$ component vanishes and the $\sigma$ components are circularly polarized, with opposite directions.
This allows us to determine the longitudinal magnetic field through measurements of circularly polarized light.
The Zeeman effect can be used to directly measure the magnetic field in stars or in laboratory plasmas.
The resulting line splitting
is $\pm0.012$\,\AA{} for the Zeeman $\pi$--$\sigma$ component separation for 5000\,\AA{} and a 1\,kG magnetic field.
The Zeeman effect is named after the Dutch physicist Pieter Zeeman, who won the Nobel prize in 1902.
Fig.~\ref{fig:sunspot} shows the effect of the magnetic field on the solar spectrum over
a sun spot.

\section{Detecting magnetic fields}
\label{sect:MF_detection}

While magnetic fields in stars can also be detected from integral light (Stokes~$I$) spectra,
if the fields are strong and the stars rotate very slowly, most of the work in this
area is done with spectropolarimeters.
The work horses for stellar magnetic field research are the low spectral resolution spectrograph
FORS\,2 at the VLT on Paranal, and the high spectral resolution spectrographs
ESPaDOnS at the CFHT on Mauna Kea,
HARPSpol at the ESO 3.6\,m on La Silla, and
NARVAL at the TBL on Pic du Midi.

Polarized light comes in two flavors, linearly and circularly polarized light.
For the magnetic field detection, we are particularly interested in the left and right circularly polarized light.
A polarimetric instrument makes use of three types of optical components.
A Wollaston prism splits linearly polarized light into ordinary and extraordinary beams.
A half wave plate rotates a polarization axis by 90$^{\circ}$.
A quarter wave plate changes linearly polarized light into circularly polarized light and back.

Different experiments measure different aspects of the magnetic field.
While the magnetic field is in every location a 3d-vector ${\cal B} = (B_{\rm x},B_{\rm y},B_{\rm z})$, 
we measure typically, averaged over the stellar disk:
\begin{itemize}
\item the longitudinal magnetic field $\left<B_{\rm z}\right>$,
\item the magnetic field modulus $\left<|B|\right>$,
\item the crossover effect $v\,\sin\,i \left<x B_{\rm z}\right>$, and
\item the mean quadratic magnetic field $\sqrt{\left<B^2\right>+\left<B_{\rm z}^2\right>}$.
\end{itemize}

These different aspects of the magnetic field are measured from different indicators
in the spectra (e.g.\ Mathys 1993).
The longitudinal magnetic field $\left<B_{\rm z}\right>$ is 
derived from measurements of wavelength shifts of spectral lines between right and left circular polarizations.
The crossover $v\,\sin\,i \left<x B_{\rm z}\right>$ is
determined from measurements of the second-order moments of line profiles in Stokes~$V$, i.e.\ the
difference of line width between opposite circular polarizations.
The mean quadratic magnetic field $\sqrt{\left<B^2\right>+\left<B_{\rm z}^2\right>}$ is
calculated from measurements of the second-order moments of line profiles in Stokes~$I$, i.e.\  the
total line widths in integral light.

\begin{figure}[t]
\centering
\includegraphics[width=0.45\textwidth, angle=270]{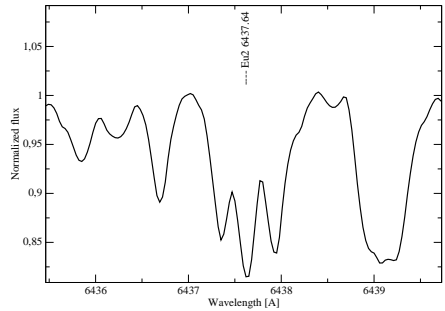}
\caption{
The spectrum of the Ap star HD\,92499 in the spectral region
around the Eu\,{\sc ii} line $\lambda\,6437.6$.
The splitting in this picture results from a magnetic field with a
mean magnetic field modulus of 8.5\,kG.
-- Credit: Fig.~4 from ``Two new chemically peculiar stars with resolved 
magnetically split lines'' by Hubrig \& Nesvacil (2007), 
MNRAS \textbf{378}, L16.
}
\label{fig:splitting}
\end{figure}

The mean magnetic field modulus $\left<|B|\right>$ is
derived from measurements of the wavelength separation of resolved magnetically split components
of spectral lines in Stokes~$I$.
E.g., Hubrig \& Nesvacil (2007) found a mean magnetic field of 
8.5\,kG in HD\,92499 (see Fig.~\ref{fig:splitting}) from 
Zeeman splitting in unpolarized light.
Johns-Krull et al.\ (1999) found a magnetic field of 
$2.6\pm0.3$\,kG in BP\,Tau from the broadening of the Ti\,{\sc i} line at 2.2233\,$\mu{}m$.

From broad-band linear polarization $(Q,U)$ a
meaningful constraint on the magnetic field can be derived from considerations
of the path followed by the star in the $(Q/I,U/I)$ plane (e.g.\ Leroy et al.\ 1996).

\section{Detecting longitudinal magnetic fields with FORS\,2}
\label{sect:MF_FORS}

\begin{figure}[t]
\centering
\includegraphics[width=0.45\textwidth]{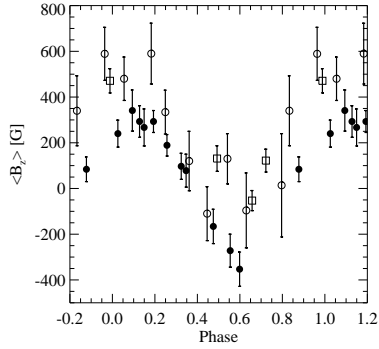}
\caption{
$\left<B_z\right>$ vs.\ the rotation phase for $\theta^1$\,Ori\,C.
{\sl Open circles:} Observations by Wade et al.\ (2006) with MuSiCoS at the TBL.
{\sl Open squares:} Observations by Petit et al.\ (2008) with ESPaDOnS and Narval.
{\sl Filled circles:} FORS\,1 measurements by Hubrig et al.\ (2008).
-- Credit: Hubrig et al., A\&A, 490, 793, 2008, reproduced with permission \copyright ESO.
}
\label{fig:theori_phasemap}
\end{figure}

With FORS\,2 in spectropolarimetric mode, we are observing the mean longitudinal magnetic field. 
The mean longitudinal magnetic field $\left<B_{\rm z}\right>$
is the component of the magnetic field parallel to the line of sight
averaged over the stellar hemisphere visible at the time of observation
and weighted by the local emergent spectral line intensity.
It depends strongly on the angles between the line of sight, the rotation axis,
and the magnetic axis, as well as the rotation phase.
Thus, it is very useful to follow stellar rotation,
but it is very limited in detecting magnetic fields from single observations.
Fig.~\ref{fig:theori_phasemap} shows measurements of the longitudinal
magnetic field of $\theta^1$\,Ori\,C.

In the oblique rotator model, we assume that we observe a magnetic field,
tilted with respect to the rotation axis, from different viewing angles, while 
the star rotates.
This magnetic field does not need to be a dipole, which can nicely be seen in the work by
Donati et al.\ (2006) on $\tau$\,Sco.

For FORS\,2 spectropolarimetric observations, we use a quarter wave plate
to go from circular polarization to linear polarization,
a half wave plate to swap the ordinary and extraordinary beams,
and a Wollaston prism to split the light with a 22$^{\prime\prime}$ beam divergence.
The spectral resolution achieved with FORS\,2 depends on the grism used, and is typically
between 2000 and 4000.

\begin{figure}[t]
\centering
\includegraphics[width=0.45\textwidth]{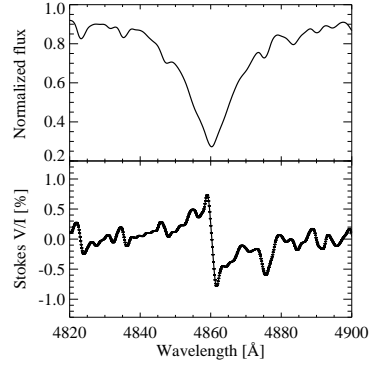}
\caption{
Stokes $I$ (top) and $V/I$ (bottom) spectra of the Ap star HD\,157751 in the vicinity of the H$\beta$ line.
In the $V/I$ spectrum, a typical Zeeman pattern can be seen, evidence for a longitudinal magnetic field
of the order of 4\,kG.
-- Credit: Fig.~2 from ``Two new chemically peculiar stars with resolved 
magnetically split lines'' by Hubrig \& Nesvacil (2007), 
MNRAS \textbf{378}, L16.
}
\label{fig:zeeman_pattern}
\end{figure}

\begin{figure}[t]
\centering
\includegraphics[width=0.45\textwidth]{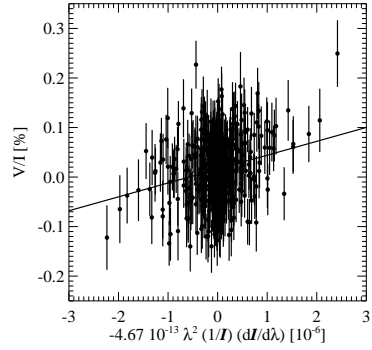}
\caption{
Linear regression detection of a $\sim$300\,G longitudinal magnetic field in
the $\beta$\,Cep star $\xi^1$\,CMa.
-- Credit: Right part of Fig.~1 from ``Discovery of magnetic fields in the {$\beta$}\,Cep star {$\xi$}$^{1}$\,CMa
and in several slowly pulsating B stars'' by Hubrig et al.\ (2006), 
MNRAS \textbf{369}, L61.
}
\label{fig:linreg}
\end{figure}
In a typical observing sequence, the retarder waveplate is set to e.g.\ $+45^{\circ}$,
an exposure is taken, which leads to two, the ordinary and the extraordinary, spectra,
then the retarder waveplate is set to $-45^{\circ}$, and another exposure is taken.
With this sequence we have now obtained four spectra.
In order to obtain a larger number of these sets of four spectra, we would continue
with further pairs of retarder waveplate settings, typical $-45^{\circ}$/$+45^{\circ}$,
$+45^{\circ}$/$-45^{\circ}$, $-45^{\circ}$/$+45^{\circ}$, etc.,
moving the retarder waveplate only every second exposure to ensure symmetry in time.
From the four or more spectra, we calculate the $V/I$ spectrum according to

\begin{equation}
\frac{V}{I} = \frac{1}{2} \left\{ \left( \frac{f^{\rm o} - f^{\rm e}}{f^{\rm o} + f^{\rm e}} \right)_{\alpha=+45^{\circ}}
                               - \left( \frac{f^{\rm o} - f^{\rm e}}{f^{\rm o} + f^{\rm e}} \right)_{\alpha=-45^{\circ}} \right\},
 \nonumber
\end{equation}

\noindent
where $\alpha$ indicates the position angle of the retarder waveplate and $f^{\rm o}$ and $f^{\rm e}$ are the ordinary and
extraordinary beams, respectively.
A typical $V/I$ spectrum for a star with a rather strong magnetic field can be seen in Fig.~\ref{fig:zeeman_pattern}.
The $V/I$ spectrum can be described by

\begin{equation} 
\frac{V}{I} = -\frac{g_{\rm eff} e \lambda^2}{4\pi{}m_ec^2}\ \frac{1}{I}\ 
\frac{{\rm d}I}{{\rm d}\lambda} \left<B_{\rm z}\right>,  \nonumber
\label{eqn:vi}
\end{equation}

\noindent
where $V$ is the Stokes parameter that measures the circular polarization, $I$
is the intensity in the unpolarized spectrum, $g_{\rm eff}$ is the effective
Land\'e factor, $e$ is the electron charge, $\lambda$ is the wavelength, $m_e$ the
electron mass, $c$ the speed of light, ${{\rm d}I/{\rm d}\lambda}$ is the
derivative of Stokes~$I$, and $\left<B_{\rm z}\right>$ is the mean longitudinal magnetic
field.
We have measured the left side of the equation, and can calculate the factor before
$\left<B_{\rm z}\right>$ from the Stokes~$I$ spectrum.
Next, we fit a linear function to the data pairs we have just generated.
The slope of this function gives the longitudinal magnetic field.
An example for such a linear regression can be seen in Fig.~\ref{fig:linreg}.

While FORS1/2 has a remarkable potential to detect magnetic fields, it is advised to
confirm these detections with high-resolution spectropolarimetry, wherever possible.
A review of FORS\,1 uncertainties can be found in Bagnulo et al.\ (2012).
Detections of a magnetic field in a number of Ap/Bp stars,
as well as pulsating B- and massive O-type stars with low-resolution spectropolarimetry
have subsequently been confirmed with high-resolution spectropolarimetry.
This was not only possible for stars exhibiting
stronger mean longitudinal magnetic fields of the order of a few hundred Gauss
(e.g.\ Hubrig et al.\ 2006, Hubrig et al.\ 2011),
but also for stars with rather weak magnetic fields of the order of 100\,G and less
(Hubrig et al.\ 2009, S\'odor et al.\ 2014).

%
%
%

%
%
%

\end{document}